\documentclass[final,times,twocolumn,sort&compress]{elsarticle}
\usepackage{amsmath}
\usepackage{amssymb}
\usepackage{graphics}
\usepackage{epstopdf}

\journal{Journal of Magnetism and Magnetic Materials}

\begin{document}

\begin{frontmatter}

\title{Effect of uniaxial single-ion anisotropy on a stability of intermediate magnetization plateaus of a spin-1 Heisenberg diamond cluster\tnoteref{grant}}  
\author[upjs]{Katar\'ina Karl'ov\'a\corref{coraut}}
\tnotetext[grant]{This work was financially supported by VEGA grant No.~1/0531/19 and by the APVV grant under the contract No. APVV-16-0186.}
\cortext[coraut]{Corresponding author}
\ead{katarina.karlova@upjs.sk}
\author[upjs]{Jozef Stre\v{c}ka} 
\author[osa]{Masayuki Hagiwara} 
\address[upjs]{Department of Theoretical Physics and Astrophysics, Faculty of Science, P. J. \v{S}af\'{a}rik University, Park Angelinum 9, 040 01 Ko\v{s}ice, Slovakia}
\address[osa]{Center for Advanced High Magnetic Field Science, Graduate School of Science, Osaka University,
Toyonaka, Osaka 560-0043, Japan}

\begin{abstract}
Ground-state phase diagrams and magnetization curves of a spin-1 Heisenberg diamond cluster with two different coupling constants and uniaxial single-ion anisotropy are investigated in a presence of the external magnetic field with the help of exact diagonalization  methods. It is shown that the spin-1 Heisenberg diamond cluster exhibits several remarkable quantum ground states, which are manifested in zero- and low-temperature magnetization curves as intermediate plateaus at one-quarter, one-half and three-quarters of the saturation magnetization. It is found that the width of the fractional magnetization plateaus depends basically on a relative strength of the coupling constants as well as uniaxial single-ion anisotropy, which may substantially shrink or even cause full breakdown of some intermediate magnetization plateaus. It is evidenced that a relatively weak uniaxial single-ion anisotropy of the easy-axis type considerably improves a theoretical fit of low-temperature magnetization curves of the tetranuclear nickel complex [Ni$_4$($\mu$-CO$_3$)$_2$(aetpy)$_8$](ClO$_4$)$_4$ in a low-field region without spoiling the previous fit based on the fully isotropic Heisenberg model in a high-field region.
\end{abstract}


\end{frontmatter}

\section{Introduction}
Molecular-based magnetic materials are considered as basic building blocks for a development of new generation of nanoscale devices with a wide application potential \cite{kahn93,siek17}. Moreover, low-temperature magnetization curves of magnetic molecules consisting from a few exchange-coupled spins may exhibit striking magnetization plateaux and jumps \cite{schn06}. In a large number of cases, the magnetizing plateaux correspond to exotic quantum states emergent due to a geometric spin frustration \cite{diep}.

In the present work, we will investigate magnetization process of a quantum spin-1 Heisenberg diamond cluster, which is motivated by the magnetic structure the homotetranuclear nickel complex [Ni$_4$($\mu$-CO$_3$)$_2$(aetpy)$_8$](ClO$_4$)$_4$ (aetpy = 2-aminoethyl-pyridine) \cite{escobar}. It has been found that the aforementioned coordination compound displays in low-temperature magnetization curves two intermediate plateaus detected at one-half and three-quarters of the saturation magnetization \cite{hagi}. The ground-state phase diagram, low-temperature magnetization curves and magnetocaloric effect of the fully isotropic spin-1 Heisenberg diamond cluster have been studied in our previous work \cite{magneto}, while in the present paper we will focus our attention to the effect of uniaxial single-ion anisotropy on the respective magnetic behaviour.

\section{Model and method}
\label{method}
\begin{figure*}
\vspace{-1.8cm}
\centering
\hspace{-0.5cm}
\includegraphics[width=0.26\textwidth]{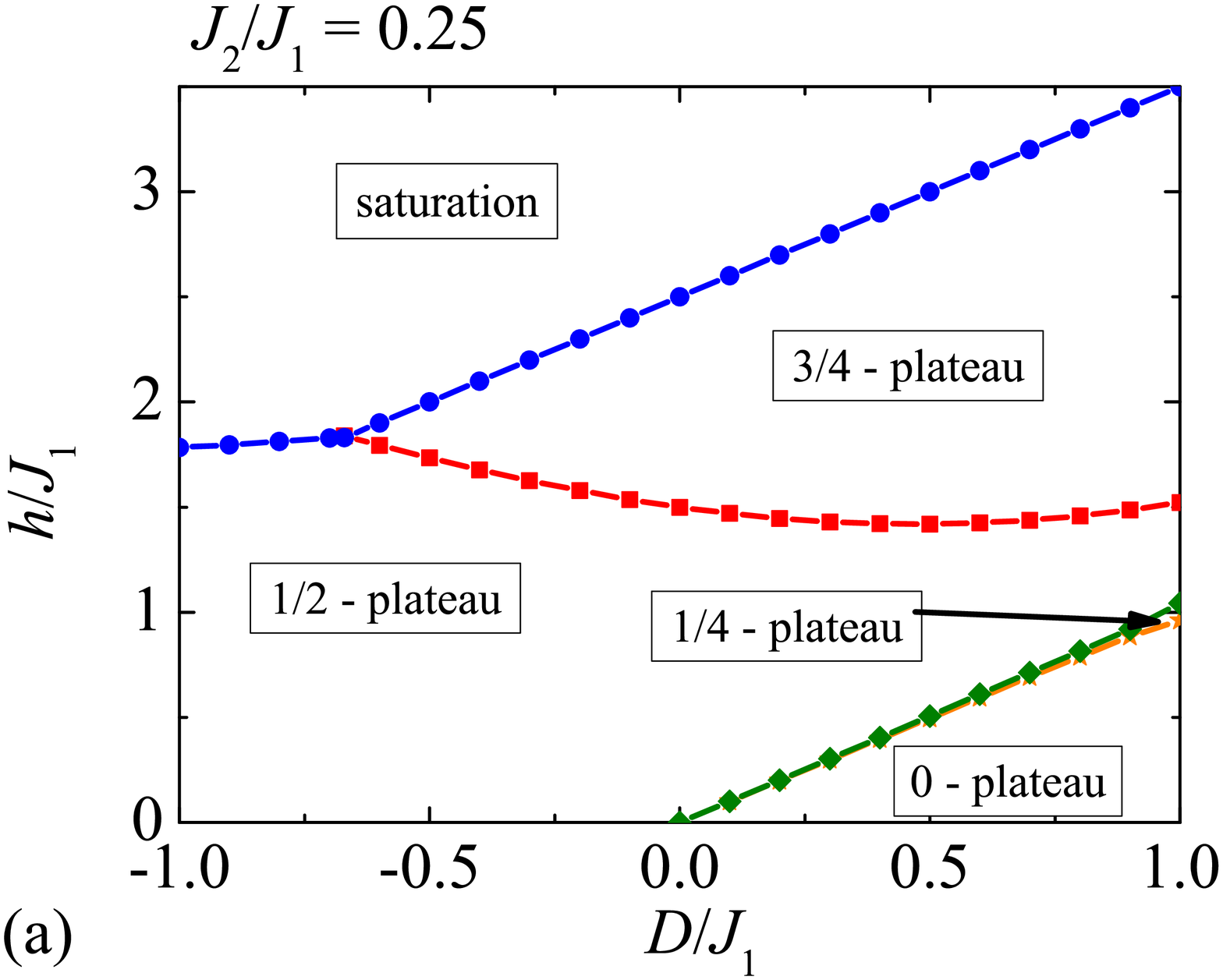}
\hspace{-0.5cm}
\includegraphics[width=0.26\textwidth]{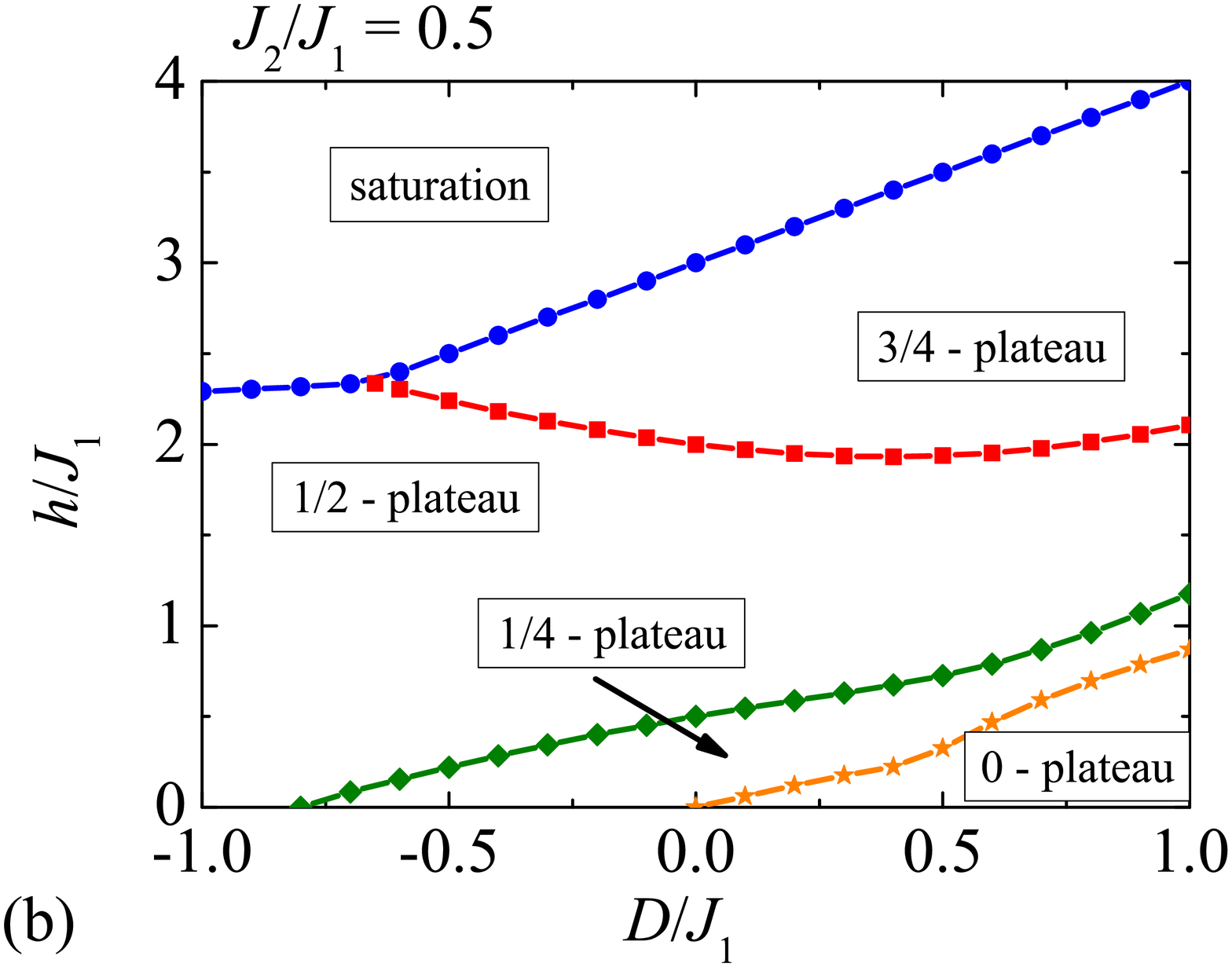}
\hspace{-0.5cm}
\includegraphics[width=0.26\textwidth]{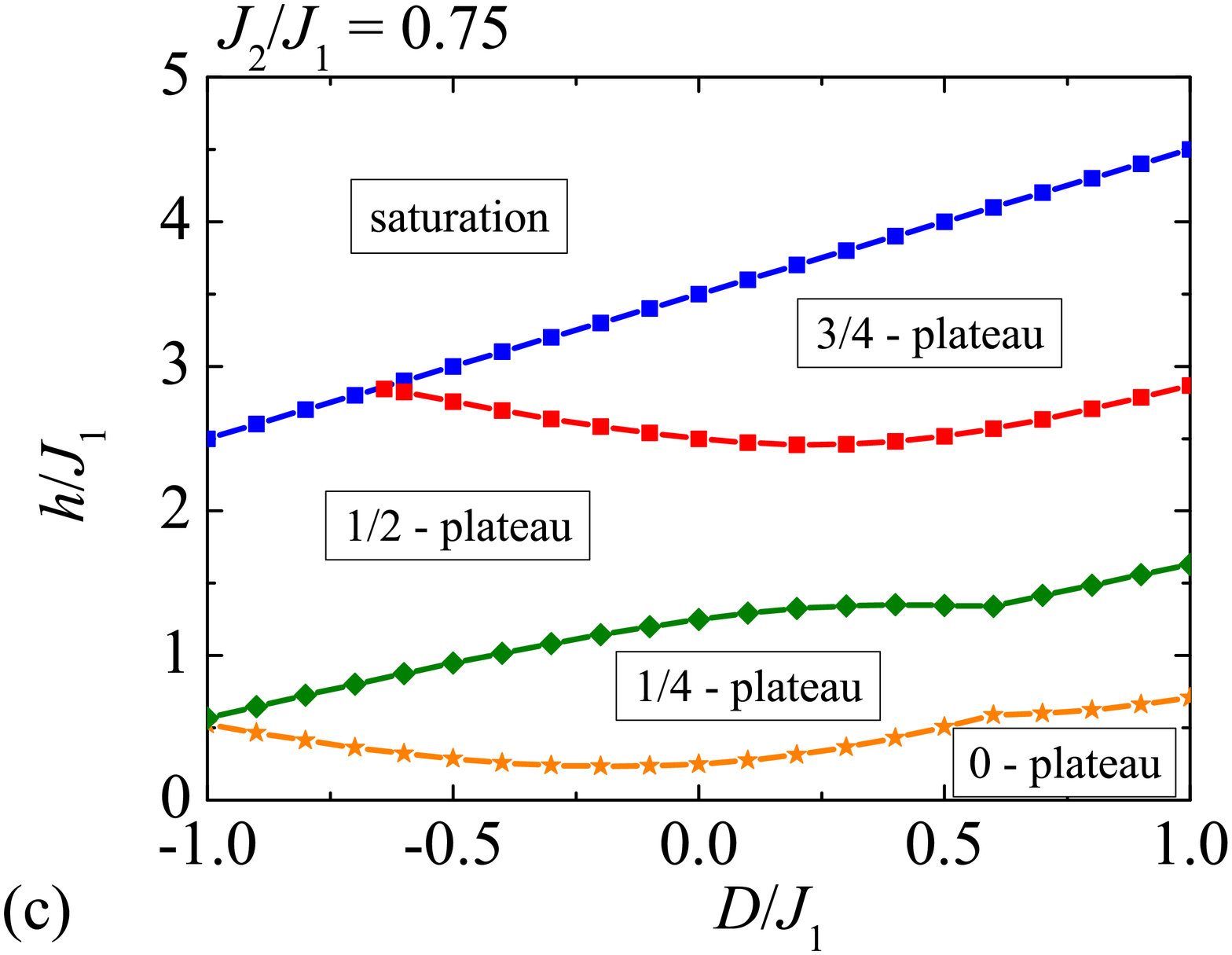}
\hspace{-0.5cm}
\includegraphics[width=0.26\textwidth]{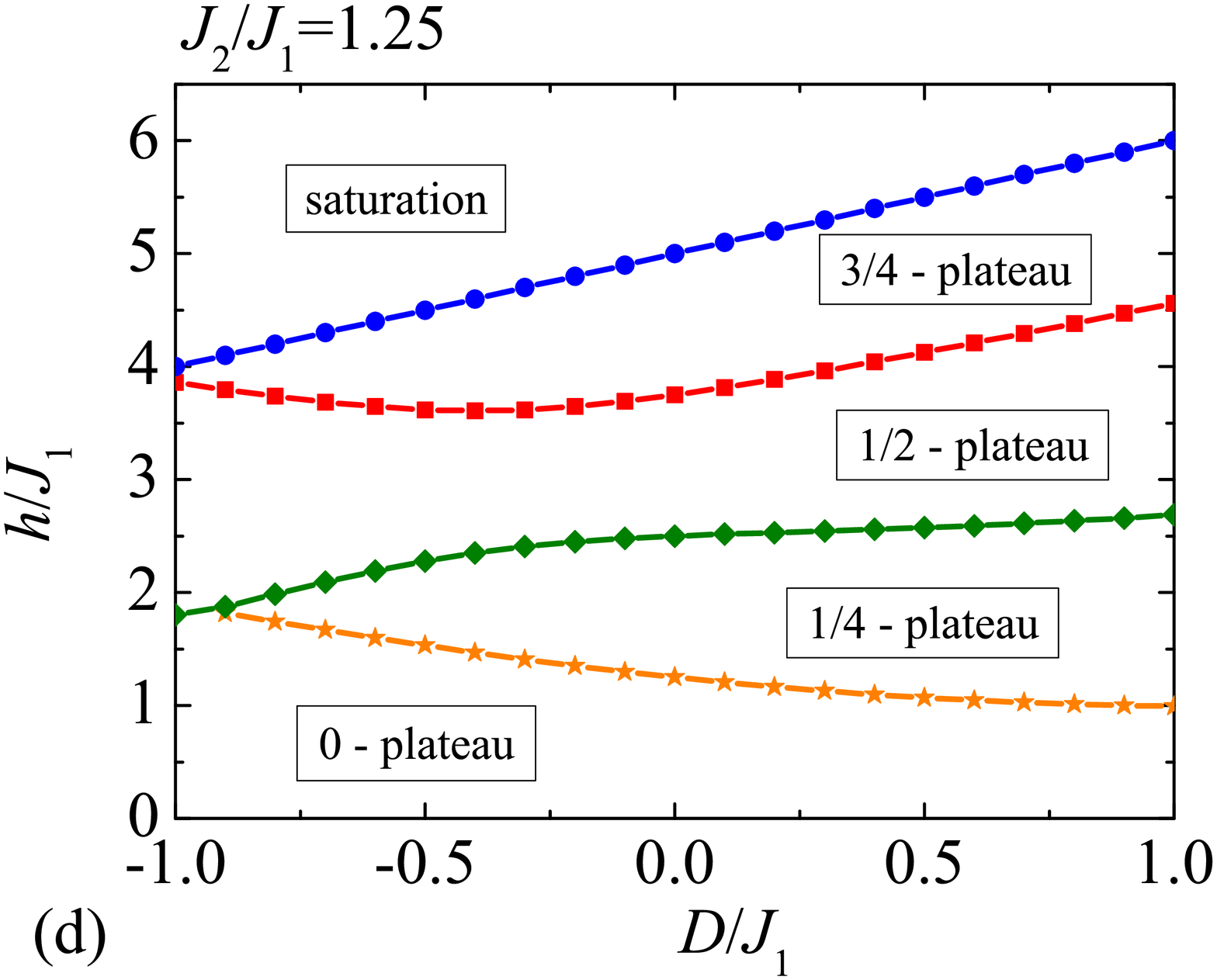}
\hspace{-0.5cm}
\vspace{-0.3cm}
\caption{The ground-state phase diagrams of the spin-1 Heisenberg diamond cluster in the $D/J_1-h/J_1$ plane for four different values of the interaction ratio: (a) $J_2/J_1=0.25$; (b) $J_2/J_1=0.5$; (c) $J_2/J_1=0.75$; (d) $J_2/J_1=1.25$.}
\label{fig1}
\end{figure*}
Let us consider the spin-1 Heisenberg diamond cluster given by the Hamiltonian
\begin{eqnarray}
	\mathbf{\hat{\mathcal{H}}} \!\!\!&=&\!\!\! J_1\mathbf{\hat{S}}_1\cdot\mathbf{\hat{S}}_2 + J_2\left(\mathbf{\hat{S}}_1 + \mathbf{\hat{S}}_2\right)\cdot\left(\mathbf{\hat{S}}_3 + \mathbf{\hat{S}}_4\right)\nonumber\\
\!\!\!&+&\!\!\!D\sum_{i=1}^4\left(\hat{S}_i^z\right)^2- h\sum_{i=1}^{4} \hat{S}_i^z,
\label{ham}
\end{eqnarray}
where $\hat{S}_{i}^{\alpha}$ $(\alpha \in \{x,y,z\}; i=1-4)$ mark spatial components of the spin-$1$ operator, $J_1$ and $J_2$ are the isotropic coupling constants for two different exchange interactions, $D$ accounts for the uniaxial single-ion anisotropy with $D<0$ ($D>0$) corresponding to the easy-axis (easy-plane) regime and the last term $h>0$ stands for the standard Zeeman's effect. The particular case without the uniaxial single-ion anisotropy ($D=0$) was comprehensively studied in our recent paper \cite{magneto}. In the present work we will therefore focus our attention to the influence of the uniaxial single-ion anisotropy upon a magnetization process. To this end, we have adapted the Lanczos method and the full exact diagonalization, which are implemented in open-source software library ALPS \cite{alps}.

\section{Results and discussion}
Let us proceed to a discussion of the most interesting results. In order to find all available ground states of the spin-1 Heisenberg diamond cluster we have implemented the numerical subroutine for the Lanczos diagonalization algorithm within the ALPS software \cite{alps}. The ground-state phase diagram of the spin-1 Heisenberg diamond cluster is plotted in Fig. \ref{fig1}(a) in the $D/J_1-h/J_1$ plane for the interaction ratio $J_2/J_1=0.25$. Five ground states are delimited by the lines of discontinuous field-induced phase transitions. It follows from Fig. \ref{fig1}(a) that the spin-1 Heisenberg diamond cluster exhibits in a full range of the considered uniaxial single-ion anisotropy $D/J_1\in(-1;1)$ relatively robust ground state manifested  in a zero-temperature magnetization curve as the 1/2-plateau, while the other robust ground state corresponding to the 3/4-plateau appears in the ground-state phase diagram whenever $D/J_1\gtrsim -0.68$ is fulfilled. On the other hand, two further ground states corresponding to the 0-plateau and narrow 1/4-plateau appear in the ground-state phase diagram only for positive values of the uniaxial single-ion anisotropy ($D>0$) of easy-plane type and they become wider with increasing of the parameter $D/J_1$.

The ground-state phase diagram of the spin-1 Heisenberg diamond cluster displayed in Fig. \ref{fig1}(b) for a higher value of the interaction ratio $J_2/J_1=0.5$ differs most significantly from the previous one through a stability region of the 1/4-plateau ground state, which becomes much wider and already appears above $D/J_1\gtrsim -0.8$. It should be noticed that a pronounced nonlinear dependence of the displayed phase boundaries on the uniaxial single-ion anisotropy is caused by nontrivial mixing of quantum states of the spin-1 Heisenberg diamond cluster. 

Next, it is worthwhile to remark that the ground-state phase diagrams of the spin-1/2 Heisenberg diamond cluster involves two ground states corresponding to the 0-plateau and 1/4-plateau in a full range of the uniaxial single-ion anisotropy $D/J_1\in(-1;1)$ if considering higher values of the interaction ratio as exemplified in Figs. \ref{fig1}(c) and (d) for two particular cases with $J_2/J_1=0.75$ and $J_2/J_1=1.25$, respectively. Moreover, it turns out that the ground state corresponding to the 3/4-plateau also appears in a full investigated range of the uniaxial single-ion anisotropy $D/J_1\in(-1;1)$ when the interaction ratio is fixed to $J_2/J_1=1.25$ [see Fig. \ref{fig1}(d)].

\begin{figure*}
\vspace{-1.8cm}
\centering
\hspace{-0.5cm}
\includegraphics[width=0.26\textwidth]{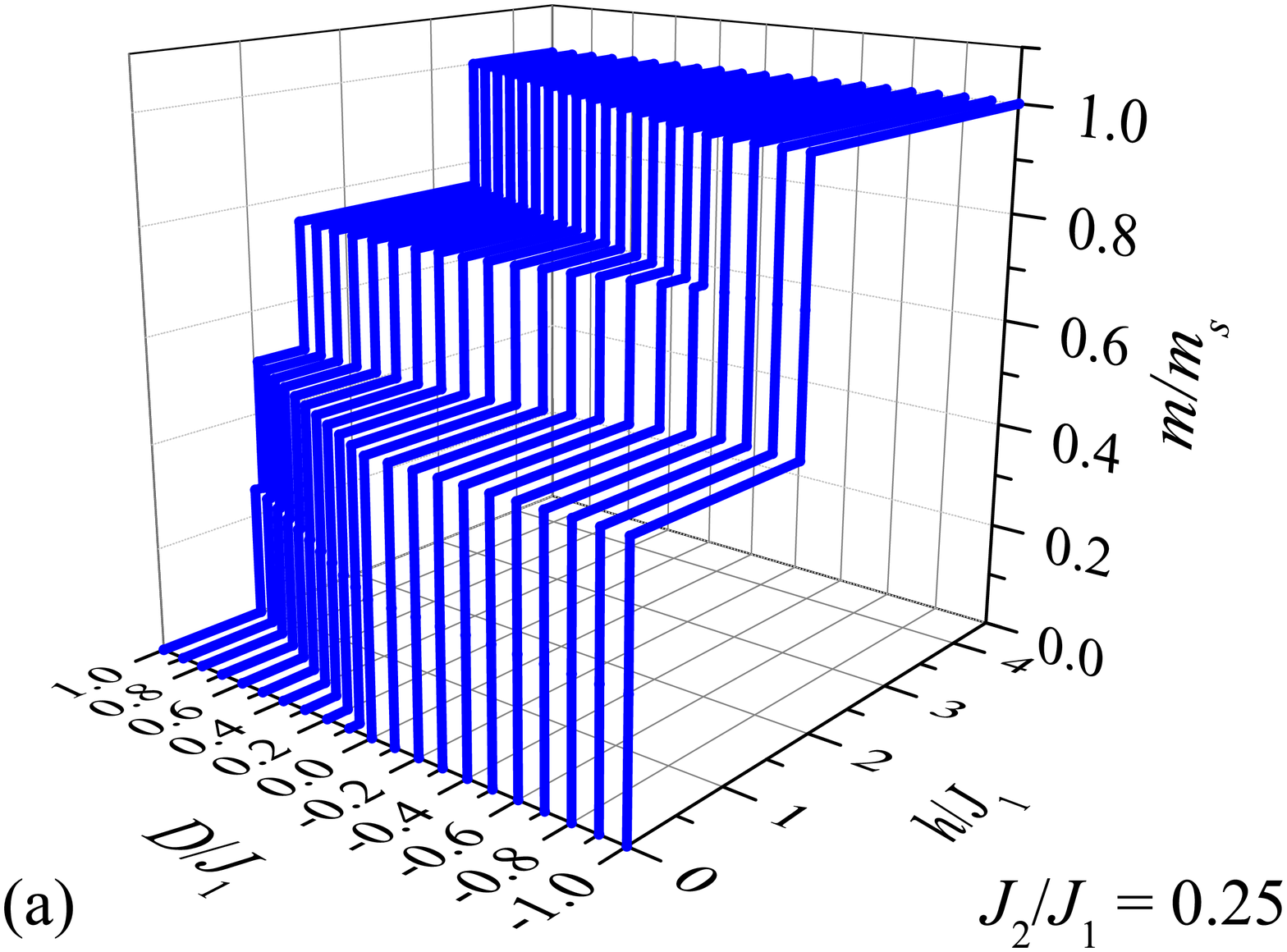}
\hspace{-0.5cm}
\includegraphics[width=0.26\textwidth]{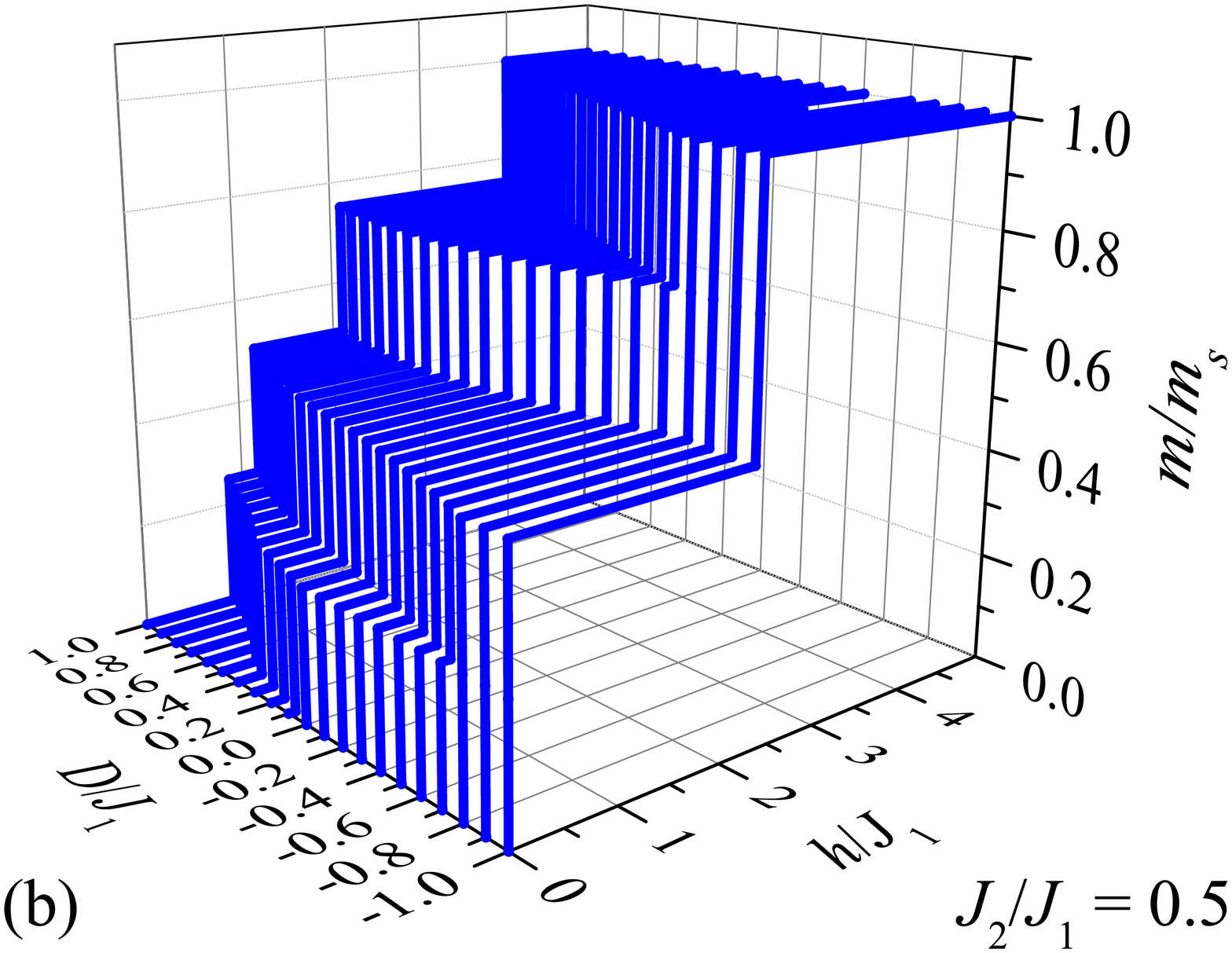}
\hspace{-0.5cm}
\includegraphics[width=0.26\textwidth]{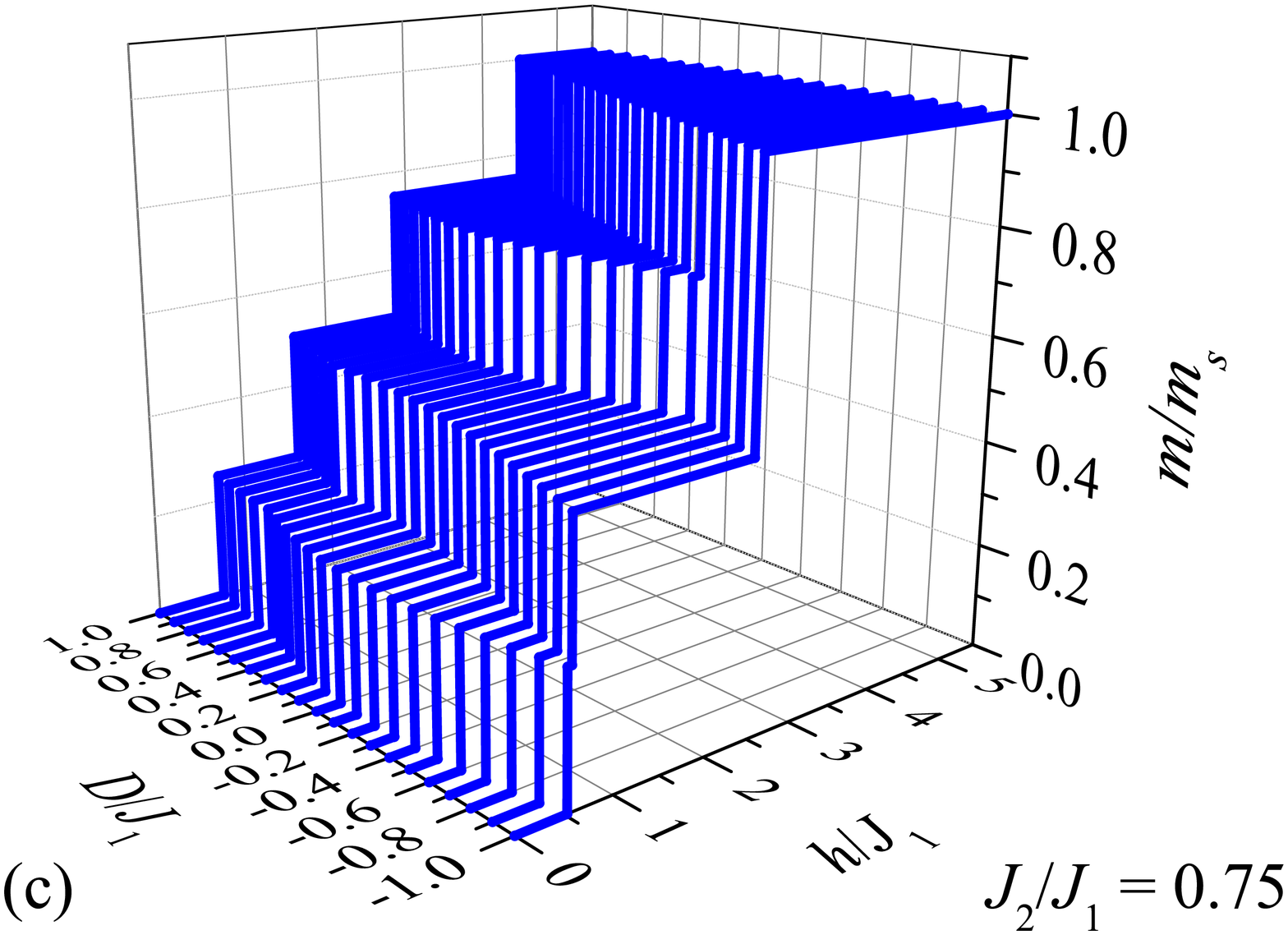}
\hspace{-0.5cm}
\includegraphics[width=0.26\textwidth]{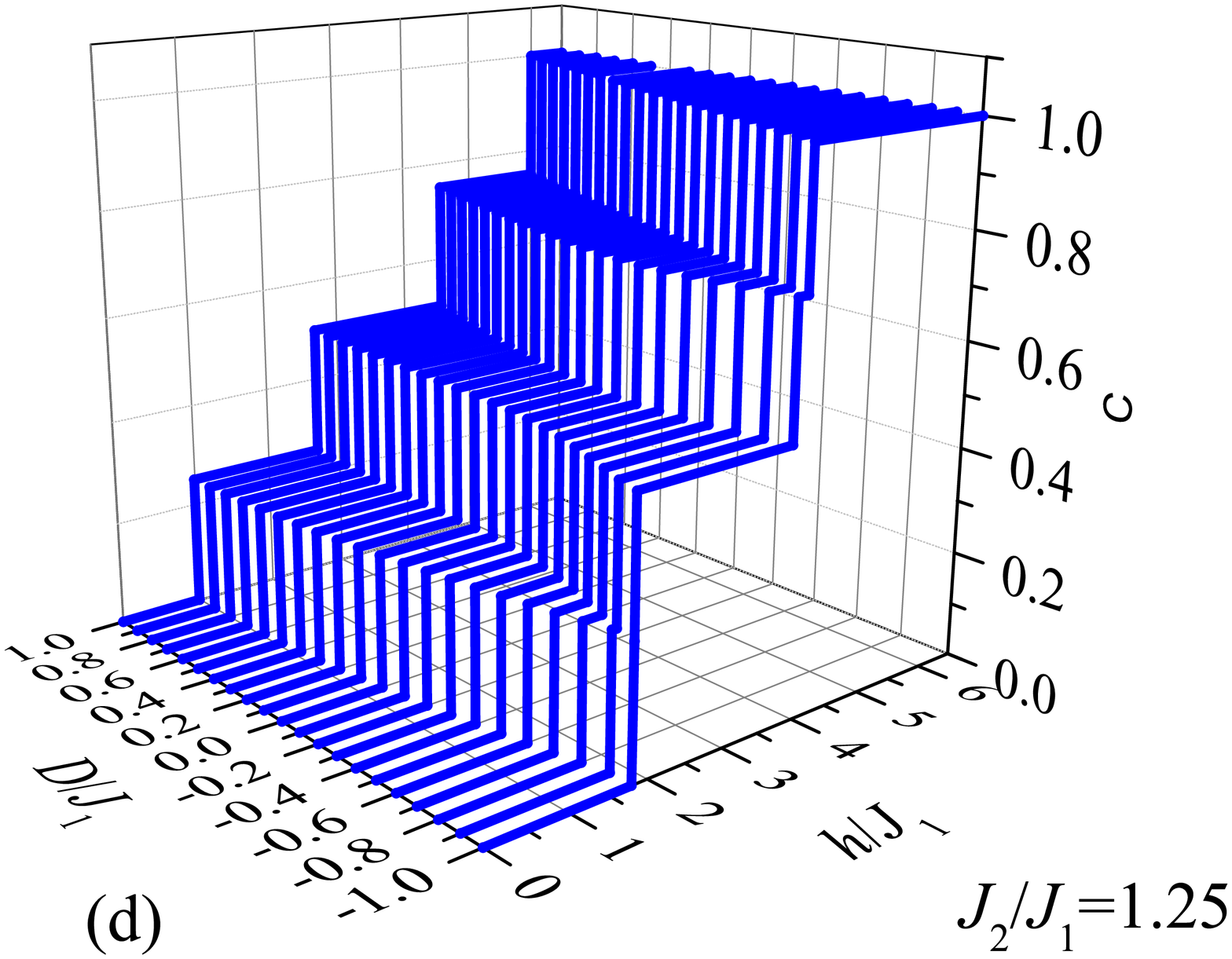}
\hspace{-0.5cm}
\vspace{-0.3cm}
\caption{Magnetization curves of the spin-1/2 Heisenberg diamond cluster as a function of magnetic field and the uniaxial single-ion anisotropy at temperature $k_{\rm B}T/J_1=0.001$ for four selected values of interaction ratio: (a) $J_2/J_1=0.25$; (b) $J_2/J_1=0.5$ (c) $J_2/J_1=0.75$; (d) $J_2/J_1=1.25$.}
\label{fig2}
\end{figure*}\begin{figure*}
\centering
\hspace{-0.5cm}
\includegraphics[width=0.26\textwidth]{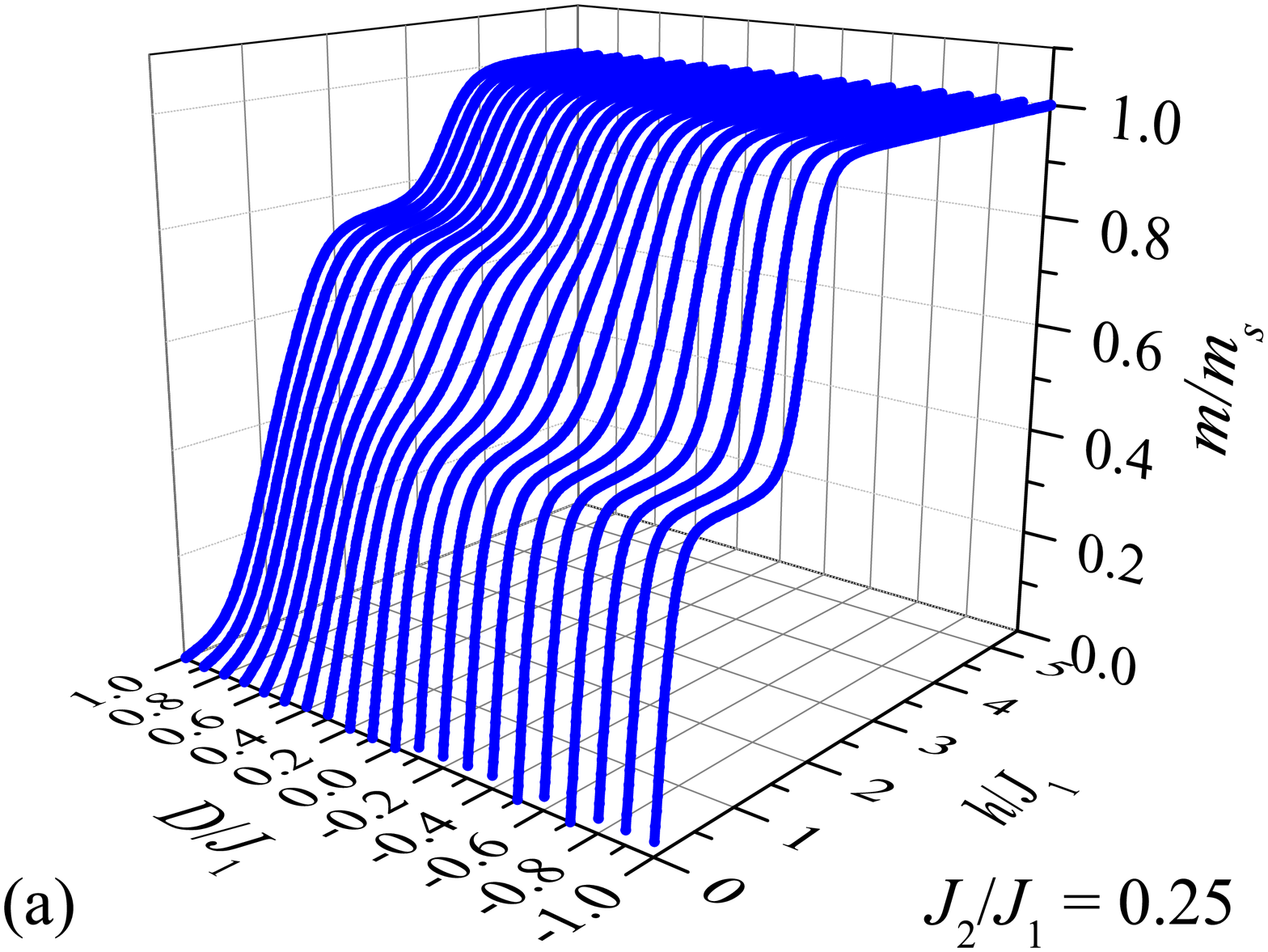}
\hspace{-0.5cm}
\includegraphics[width=0.26\textwidth]{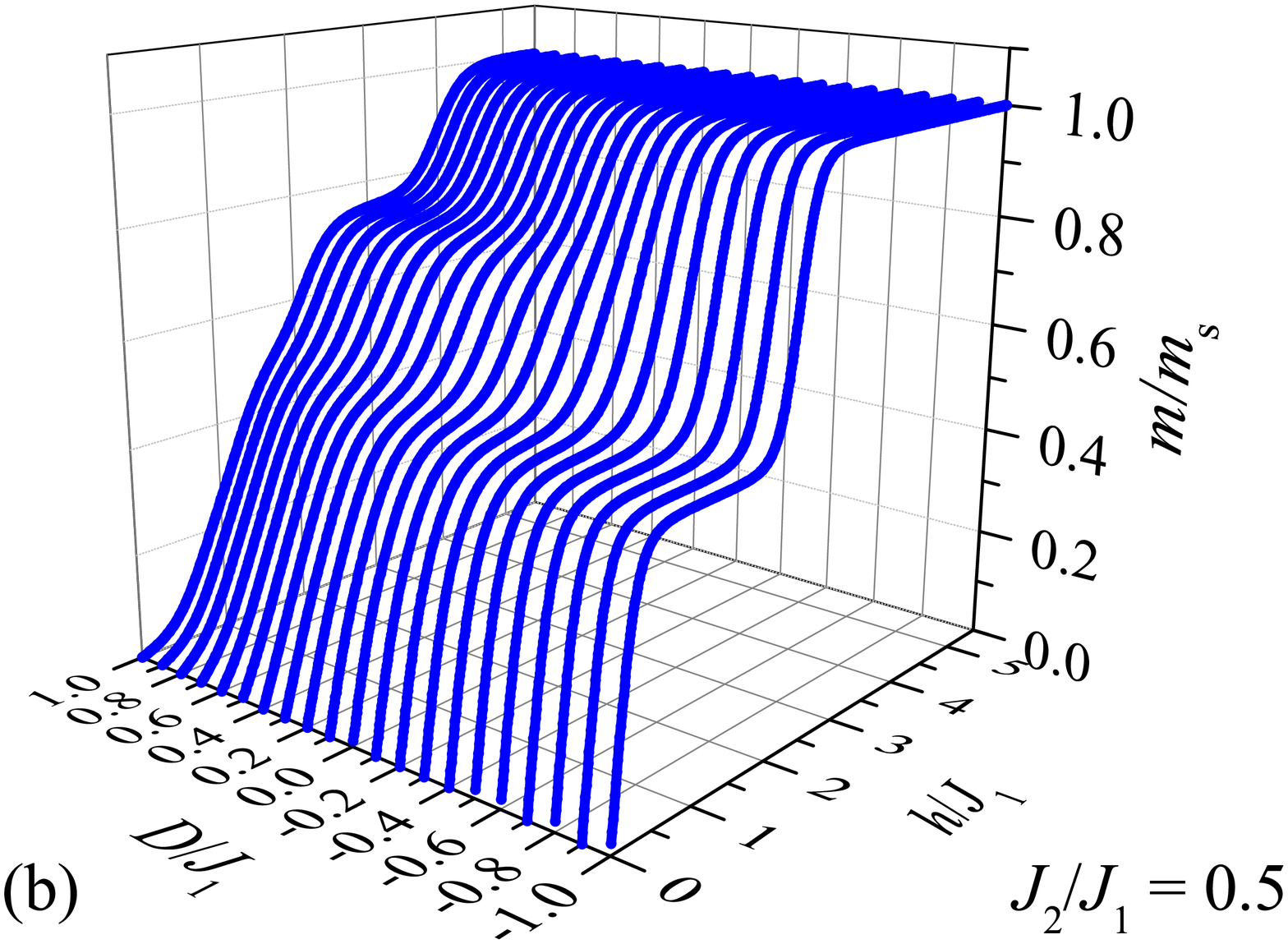}
\hspace{-0.5cm}
\includegraphics[width=0.26\textwidth]{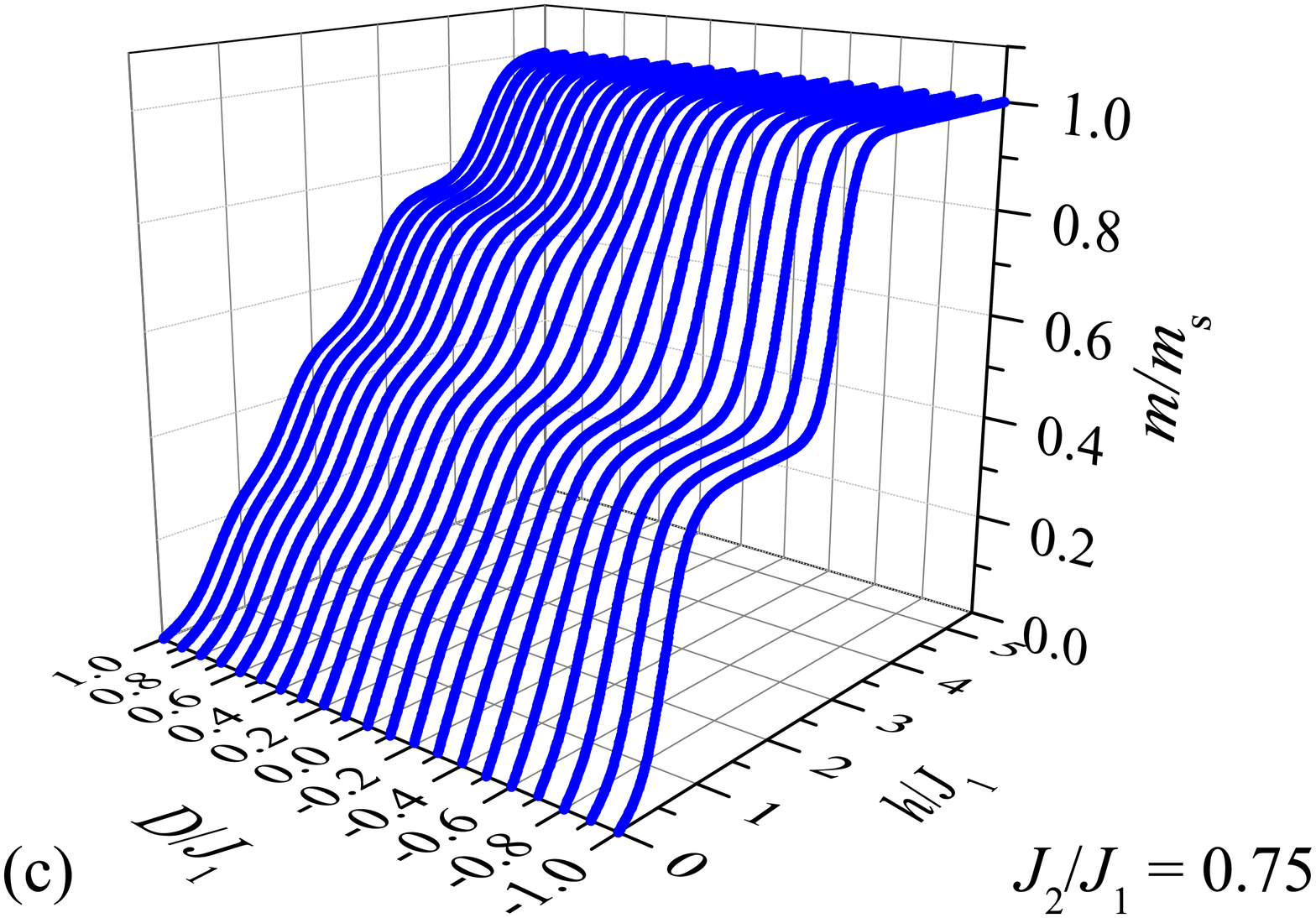}
\hspace{-0.5cm}
\includegraphics[width=0.26\textwidth]{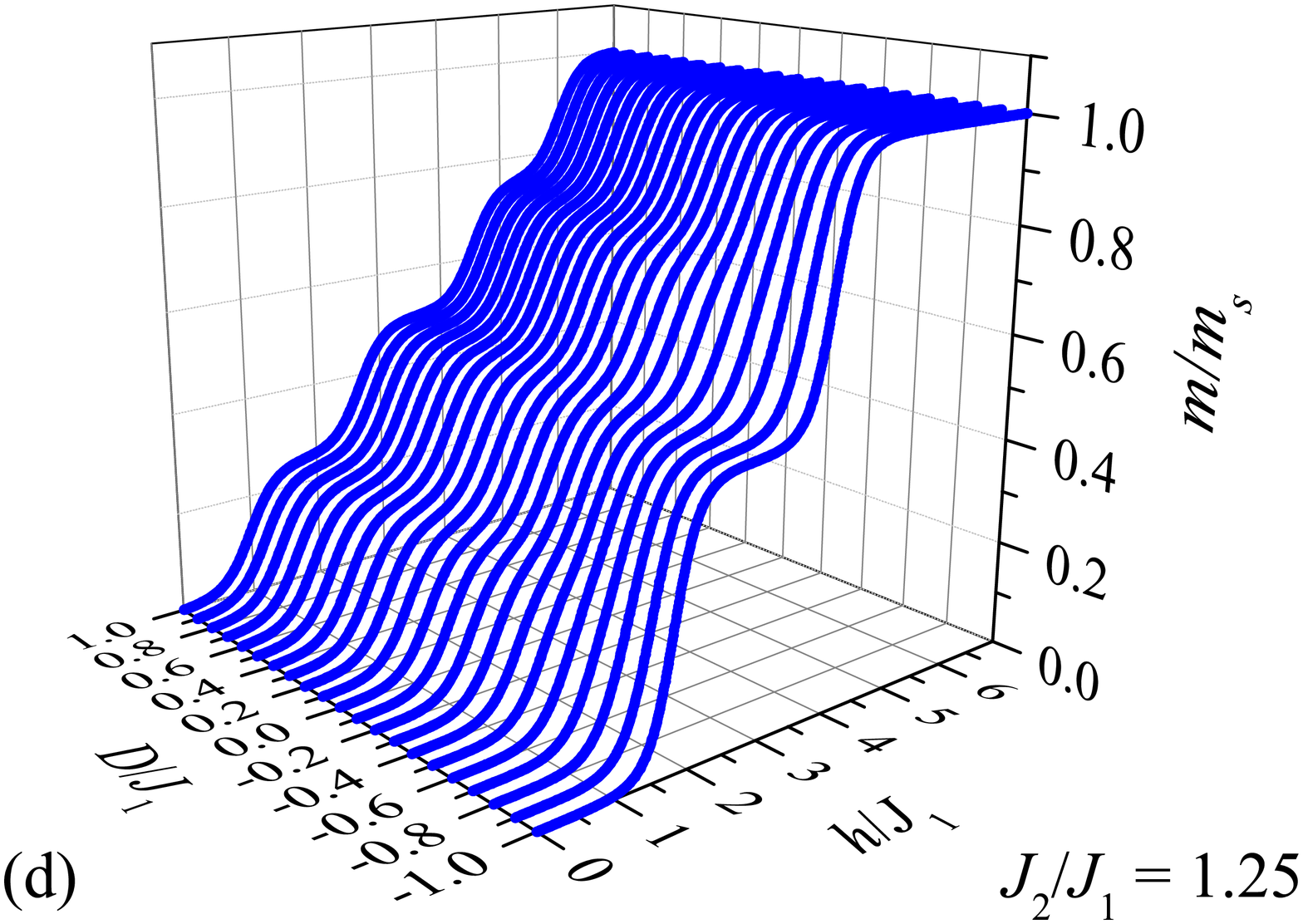}
\hspace{-0.5cm}
\vspace{-0.3cm}
\caption{Magnetization curves of the spin-1/2 Heisenberg diamond cluster as a function of magnetic field and the uniaxial single-ion anisotropy at temperature $k_{\rm B}T/J_1=0.2$ for four selected values of the interaction ratio: (a) $J_2/J_1=0.25$; (b) $J_2/J_1=0.5$ (c) $J_2/J_1=0.75$; (d) $J_2/J_1=1.25$.}
\label{fig3}
\end{figure*}

To bring insight into how temperature influences the magnetization process of the  spin-1 Heisenberg diamond cluster, the isothermal magnetization curves at sufficiently low temperature $k_{\rm B}T/J_1=0.001$ computed using the full exact diagonalization within ALPS project \cite{alps} are depicted in Fig. \ref{fig2}. As one can see, the magnetization process at very low temperature are in perfect agreement with the established ground-state phase diagrams with exception of discontinuous magnetization jumps, which turn to a steep but continuous changes of the magnetization at any finite temperature. However, this fact cannot be seen by eye when temperature is as low as $k_{\rm B}T/J_1=0.001$. In any case, the results obtained from the full exact diagonalization at very low temperature thus provide an independent confirmation of the ground-state phase diagrams obtained from the Lanczos method. Indeed, the 1/2-plateau emerges in a magnetization process in a full range of the uniaxial single-ion anisotropies $D/J_1\in(-1;1)$ for all selected values of interaction ratio $J_2/J_1=0.25, 0.5, 0.75$ and 1.25, while the 3/4-plateau is absent in a magnetization process of the spin-1 Heisenberg diamond cluster for strong negative values of the uniaxial single-ion anisotropy $D/J_1\lesssim-0.68$ and three selected values of the interaction ratio $J_2/J_1=0.25, 0.5 $ and 0.75 [see Figs. \ref{fig2}(a)-(c)]. Furthermore, the 0-plateau can be observed in  a full range of the uniaxial single-ion anisotropy  $D/J_1\in(-1;1)$ for $J_2/J_1=0.75$ and 1.25 [see Figs. \ref{fig2}(c)-(d)], while presence of the 1/4-plateau crucially depends on a mutual interplay between the uniaxial single-ion anisotropy and the interaction ratio. The 1/4-plateau can be nevertheless detected in a full range of the uniaxial single-ion anisotropy $D/J_1\in(-1;1)$ if the interaction ratio is fixed to $J_2/J_1=0.75$. 

The isothermal magnetization curves of the spin-1 Heisenberg diamond cluster at moderate temperature $k_{\rm B}T/J_1=0.2$  are shown in Fig. \ref{fig3} as a function of the magnetic field for several values of the uniaxial single-ion anisotropy and four selected values of the interaction ratio $J_2/J_1$. As one can see, only relatively robust magnetization plateaux are still discernible at moderate temperature, mainly the 1/2- and 3/4-plateau [see Figs. \ref{fig3}(a)-(c)], while other tiny intermediate magnetization plateaux are clearly discernible only when the interaction ratio $J_2/J_1=1.25$ is fixed and the uniaxial single-ion anisotropy of the easy-plane type $D/J_1\gg 0$ is sufficiently strong [see Fig. \ref{fig3}(d)].

\begin{figure}[!thb]
\centering
\includegraphics[width=0.4\textwidth]{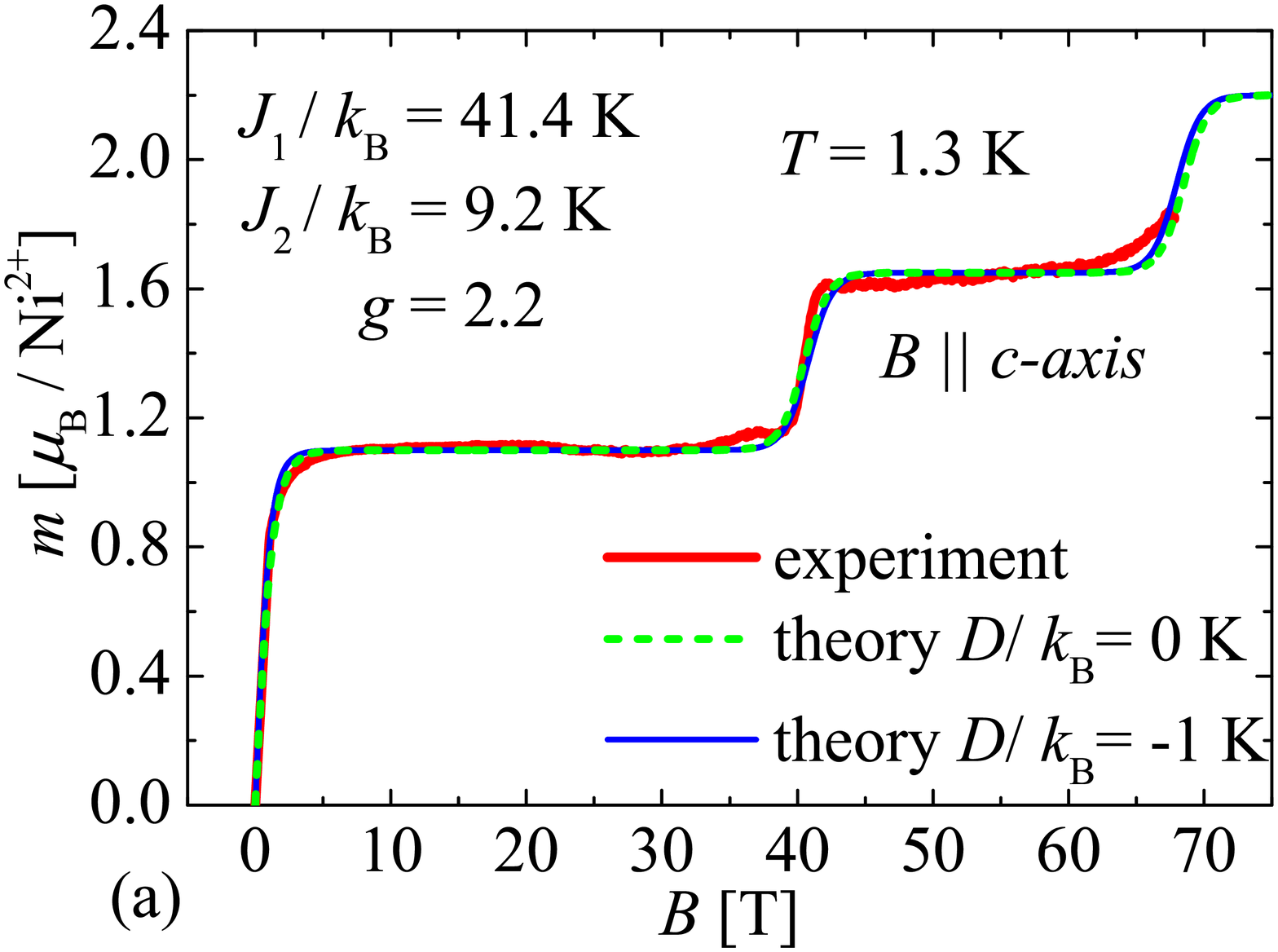}
\includegraphics[width=0.4\textwidth]{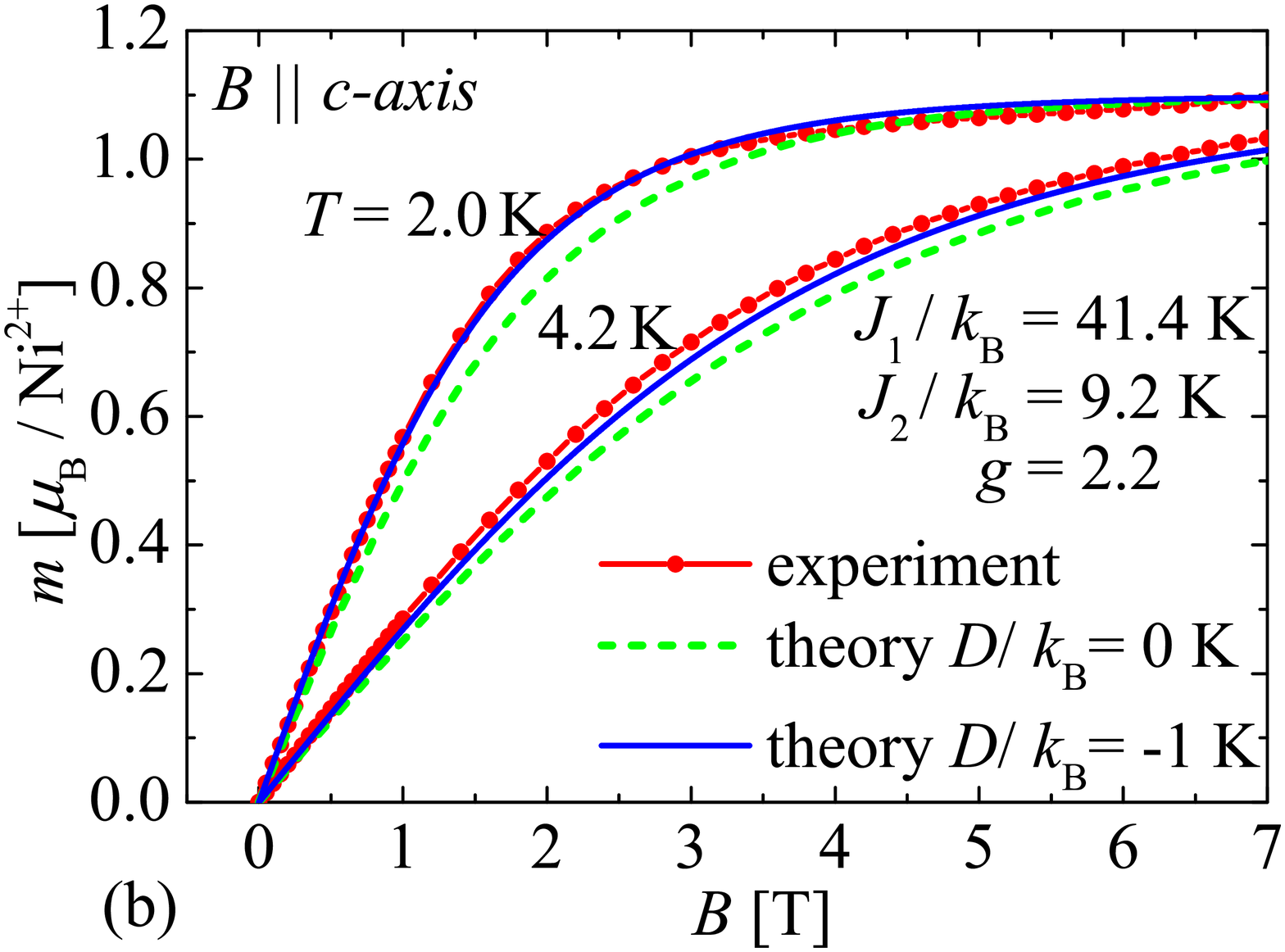}
\vspace{-0.3cm}
\caption{(a) Magnetization curve of the tetranuclear nickel compound [Ni$_4$($\mu$-CO$_3$)$_2$(aetpy)$_8$](ClO$_4$)$_4$ (red thick line) measured in pulsed magnetic fields up to 68 T at sufficiently low temperature 1.3 K and the corresponding theoretical fit based on the spin-1 Heisenberg diamond cluster with (without) the uniaxial single-ion anisotropy $D/k_{\rm B}=-1 {\rm{K}}$ ($D/k_{\rm B}=0 {\rm{K}}$) shown as a blue solid (green dashed) line; (b) Magnetization curves (red lines with filled circles) of the tetranuclear nickel compound [Ni$_4$($\mu$-CO$_3$)$_2$(aetpy)$_8$](ClO$_4$)$_4$ in static magnetic fields up to 7 T at two different temperatures 2.0 K and 4.2 K versus the respective theoretical fit based on the spin-1 Heisenberg diamond cluster with (without) the uniaxial single-ion anisotropy $D/k_{\rm B}=-1 {\rm{K}}$ ($D/k_{\rm B}=0 {\rm{K}}$) shown as a solid blue (green dashed) line. According to Ref. \cite{magneto}, other fitting parameters are fixed to: $J_1/k_{\rm B}$ = 41.4 K, $J_2/k_{\rm B}$ = 9.2 K, $g$ = 2.2.}
\label{fig4}
\end{figure}

High-field magnetization data of the tetranuclear nickel complex [Ni$_4$($\mu$-CO$_3$)$_2$(aetpy)$_8$](ClO$_4$)$_4$ measured in pulsed magnetic fields up to 68 T at sufficiently low temperature 1.3 K are compared in Fig. \ref{fig4}(a) with the respective theoretical results for the spin-1 Heisenberg diamond cluster with or without the uniaxial single-ion anisotropy. The best fitting set reported previously for the theoretical model without the uniaxial single-ion anisotropy $J_1/k_{\rm B}$ = 41.4 K, $J_2/k_{\rm B}$ = 9.2 K and $g$ = 2.2 was justified by matching the respective theoretical results with experimentally observed magnetization jumps related to a breakdown of the intermediate 1/2- and 3/4-plateaux \cite{magneto}. It is noteworthy that a strength of the uniaxial single-ion anisotropy must be relatively small, if any, because both experimentally observed magnetic fields related to the magnetization jumps are almost independent of a spatial orientation of the applied magnetic field \cite{hagi}.

In order not to spoil the reasonable fit of the magnetization curve in a high-field region we have decided to keep the same fitting set of the parameters \cite{magneto} under the concurrent effect of a weak uniaxial single-ion anisotropy. It turned out that a weak easy-axis uniaxial single-ion anisotropy substantially improves a theoretical fit of low-field magnetization data of the tetranuclear nickel complex [Ni$_4$($\mu$-CO$_3$)$_2$(aetpy)$_8$](ClO$_4$)$_4$, which were previously recorded in static magnetic fields up to 7 T at low enough temperatures 2.0 K and 4.2 K \cite{hagi}. It is evident from Fig. \ref{fig4}(b) that the theoretical prediction based on the spin-1 Heisenberg diamond cluster with a weak easy-axis uniaxial single-ion anisotropy $D/k_{\rm B}=-1~{\rm{K}}$ fits much better the measured low-field magnetization data than that without the uniaxial single-ion anisotropy.
  
\section{Conclusion}
\label{conclusion}
The present paper deals with the ground-state phase diagram and magnetization curves of the spin-1 Heisenberg diamond cluster, which accounts for the uniaxial single-ion anisotropy. It has been verified that the spin-1 Heisenberg diamond cluster may exhibit intermediate magnetization plateaux at 0, 1/4, 1/2 and 3/4 of the saturation magnetization quite similarly as its fully isotropic counterpart \cite{magneto}, whereby the origin and termination of intermediate plateaux is basically affected by a mutual interplay between the interaction ratio and the uniaxial single-ion anisotropy. Magnetization curves are strongly smeared out by rising temperature and only sizable magnetization plateaux are still discernible at moderate temperatures. Last but not least, it has been evidenced that a weak easy-axis uniaxial single-ion anisotropy basically improves a theoretical fit of low-field magnetization curves for the nickel-based compound [Ni$_4$($\mu$-CO$_3$)$_2$(aetpy)$_8$](ClO$_4$)$_4$.

\end{document}